\documentclass[aps,prl, twocolumn,groupedaddress,showpacs]{revtex4} 

\usepackage{graphicx}
\usepackage{epstopdf}
\usepackage{amssymb}  

\def\fwhm {full-width half-maximum}
\def\mw {$\lambda_{m}$}
\def\lw {$\lambda_{L}$}
\def\MOT {magneto-optical trap}

\newcommand{\degrees}{$^{\circ}$}
\newcommand{\degC}{$^{\circ}$C}

\newcommand{\si}{$\sim$}
\newcommand{\microns}{$\mu$m}
\newcommand{\uK}{$\mu$K}
\newcommand{\uW}{$\mu$W}

\newcommand{\ee}[1]{\ensuremath{\times 10^{ #1}}}

\newcommand{\coolingT}{$^{1}S_{0}\leftrightarrow\,^{3}P_{1}$}  

\newcommand{\clockTboth}{$^{1}S_{0} \leftrightarrow\,^{3}P_{0}$} 

\newcommand{\Hg}{$^{199}$Hg}

\begin{document}

\title{A neutral atom frequency reference in the deep UV with $10^{-15}$ range uncertainty}  %

\author{J.~J.~McFerran,  L.~Yi,  S.~Mejri,   S. Di Manno, W.~Zhang,  J.~Gu\'{e}na, Y.~Le~Coq,  and S.~Bize} 
\email[]{sebastien.bize@obspm.fr}

\affiliation{LNE-SYRTE, Observatoire de Paris, CNRS, UPMC, 61 Avenue de l'Observatoire, 75014, Paris, France} 

\date{\today}

\begin{abstract}
We present an assessment of the (6$s^{2}$) $^{1}S_{0} \leftrightarrow$ ($6s7s$) $^{3}P_{0}$ clock transition frequency in $^{199}$Hg with an uncertainty reduction of nearly  three orders of magnitude and demonstrate an atomic quality factor, $Q$,  of \si$10^{14}$. The \Hg\ atoms are confined in a vertical lattice trap with light at the newly determined magic wavelength of $362.5697\pm0.0011$\,nm  and at a lattice depth of 20\,$E_{R}$. The atoms are loaded from a single stage \MOT\ with cooling light at 253.7\,nm. The high $Q$  factor is obtained with an 80\,ms Rabi pulse  at  265.6\,nm. The frequency of the clock transition is found to be 1 128 575 290 808 162.0 $\pm$ 6.4 (sys.) $\pm$ 0.3 (stat.)\,Hz (fractional uncertainty = 5.7\ee{-15}). Neither an atom number nor  second order Zeeman dependence have yet to be detected. Only three laser wavelengths are used for the cooling, lattice trapping, probing and detection.

\end{abstract} 

\pacs{
 32.30.Jc 
 37.10.Jk 
 32.10.Dk 
 42.62.Fi 
 37.10.De 
 }


\maketitle



The advance in performance of atomic clocks over recent decades  is an impressive  accomplishment. The  accuracy  of microwave clocks has improved at a rate of \si10 per decade and now optical clocks are showing an improvement at a rate  $>10^{2}$ per decade~\cite{Mar2010}. Optical clock technology is still relatively young; yet four atomic species have demonstrated line-center frequency uncertainties at or below $10^{-16}$~\cite{Lud2008, Ros2008, Cho2010}, with the potential of several more to follow~\cite{Hac2008,Yi2011,Tam2009,Yud2011}. This will enable further precision in tests searching for new physics in the low energy regime~\cite{Bla2008, For2007, Kar2008}.

State of the art optical clocks demand an exceptionally high line-$Q$, since  the attainable frequency stability of a local oscillator locked to an atomic transition is inversely proportional to this quality factor. As $Q= \nu/\Delta\nu$, where   $\nu$ is the  clock transition frequency  and $\Delta\nu$  is the excitation linewidth, it is advantageous to have a carrier frequency as high as practicality allows.  In a neutral mercury clock this frequency is  approximately 1.128\,PHz, lying in the deep ultraviolet, hence favourable in this respect. 

Ion clocks have demonstrated impressive results with respect to line frequency uncertainties~\cite{Tam2007,Ros2008, Cho2010a} and progress is expected to continue. A prime driver for neutral atom clocks is the significantly higher number of quantum absorbers used for the atom$-$probe interaction. This provides an additional measurement lever by reducing the integration time required for assessing various clock shifts. The tightly bound atoms in an optical lattice trap become heavily immune to Doppler and photon-recoil effects (Lamb-Dicke regime). In the process of constraining the atoms, one needs to shift the upper and lower clock state energies by equal amounts by maintaining the lattice light at the magic wavelength~\cite{Kat2003}. We previously reported the measurement of the magic wavelength in $^{199}$Hg with 0.21\,nm uncertainty~\cite{Yi2011}, making it the third element to be tested as an optical lattice clock. Like other alkaline-earth-metal type elements mercury has a doubly forbidden transition between the ground (n$s^{2}$)  $^{1}S_{0}$  and the excited (n$s$n$p$) $^{3}P_{0}$ levels ($n=6$ for Hg). It  has several favourable characteristics for its use as a primary frequency reference; the most significant being that  its sensitivity to blackbody radiation (BBR)  is more than an order of magnitude lower than that of  Sr and Yb~\cite{Por2006,Hac2008}.  The uncertainty due to the BBR dominates the frequency uncertainty budgets of these two clocks~\cite{Swa2010,Lem2009a}.  Furthermore, the absence of a high temperature oven for Hg helps to reduce the temperature variations in the vicinity of the trapped atoms, thus helping to reduce the BBR uncertainty further. 

Initial clock transition spectral widths with $^{199}$Hg were about 2\,kHz due to delocalization of atomic states across lattice sites~\cite{Yi2011, Mej2011}. Here we have increased the lattice depth to 20 $E_{R}$ (recoil energy = $\hbar^{2} k_{l}^{2}/2m$; $k_{l}$ is the  wave number; $m$ is the atomic mass) forming more tightly bound atoms. This has enabled a linewidth reduction to \si150\,Hz from which a series of light shift measurements permitted an improved estimate of the magic wavelength. In combination with Zeeman shift measurements this has led to further narrowing, reaching 10\,Hz.  In this letter we report results pertaining to both the 150\,Hz  and \si10\,Hz wide lines; highlighting the difference where appropriate. Our narrowest lines correspond to a quality factor of $Q=10^{14}$, which has only previously been obtained in a limited number of systems; e.g., ~\cite{Raf2000, Boy2007, Lem2009a, Cho2011, Hun2011}. We have conducted measurements related to various systematic shifts, including the first order light shift, Zeeman shifts and atom number dependence.

\begin{figure}[h]
\begin{center}
{		
\includegraphics[width=8.8cm,keepaspectratio=true]{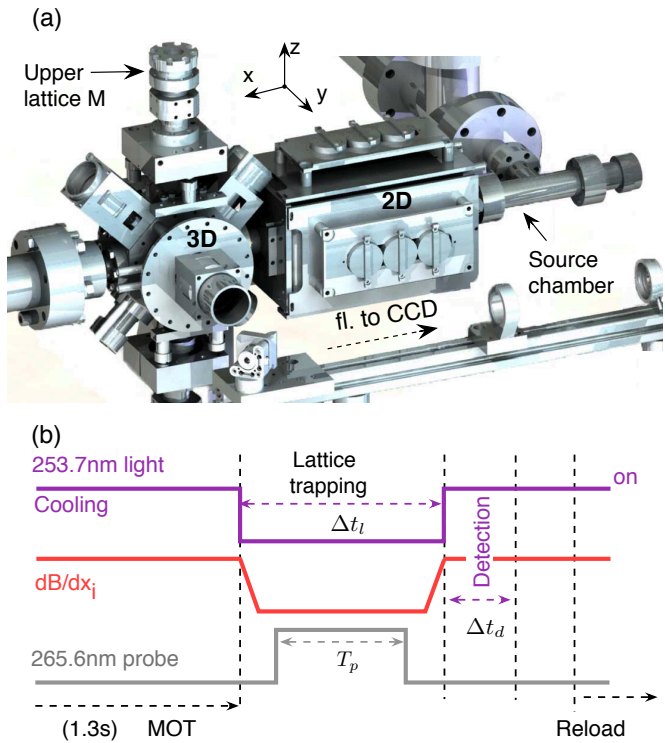}}
\caption[]{\footnotesize (Color online)  (a) Drawing of the vacuum system showing the 3D MOT (left side) and 2D MOT chambers. Also visible is the upper mirror of the optical lattice cavity above the 3D MOT chamber; fl., fluoresence. (b) Timing sequence for the cooling light, probe light and 3D MOT magnetic field gradient. The lattice light remains on continuously throughout the cycle.  $\Delta t_{l}=80$\,ms and $T_{p}=50$\,ms for most measurements here; $\Delta t_{d}=9$\,ms.
} \label{ChamberandSequence}		
\end{center}
\end{figure}

The \Hg\ atoms are loaded into a 1D vertical optical lattice from a single-stage \MOT\ (MOT) that uses laser light at 253.7\,nm for the cooling  (\coolingT). The MOT and lattice trap are described in detail in \cite{Mej2011}. Some modifications have been made to the lattice cavity and to the resonant cavity generating the 253.7\,nm ultraviolet light, which we outline here. A scaled drawing of the apparatus is shown in Fig.~\ref{ChamberandSequence}(a). On the left hand side appears the combined MOT chamber and optical lattice cavity. The ports in the horizontal and 45\degrees\ directions provide access for the MOT beams (from beneath in the case of the 45\degrees\ ports). On the right hand side is the mercury source chamber, which includes a 2D MOT set-up. The 2D MOT is not employed in the measurements described here, but its use is anticipated in future.   Several grams of mercury are maintained at about $-40$\,\degC\ using a dual Peltier stage. Between 50 and 70\,mW of 253.7\,nm light is generated from a frequency quadrupling scheme for use in the 3D MOT (with the variation in power occurring over weeks or months).

The vertically orientated lattice cavity is comprised of two spherical mirrors, both with a radius of curvature of 250\,mm and has a finesse of 210 at 362\,nm.  The waist size (rad.) is 120\,\microns\ where the lattice light overlaps the MOT cloud and produces a maximum lattice depth of 25\,$E_{R}$ (or 9.2\,\uK).  Light at or near the magic wavelength is coupled into the cavity from below and the transmitted light is used to form a side-lock to  maintain constant intensity.  A tunable Ti:sapphire laser has its output frequency doubled in a LiB$_{3}$O$_{5}$  crystal based resonant cavity  to produce the lattice light.  In our previous lattice cavity design there was a 45\degrees\ reflector that formed an L-shaped cavity to lift polarization degeneracy~\cite{Mej2011}.  This optic was found to degrade rather quickly under vacuum with the incidence of 362\,nm light at high power and has been removed.

\begin{figure}[h]
\begin{center}
{		
\includegraphics[width=8.6cm,keepaspectratio=true]{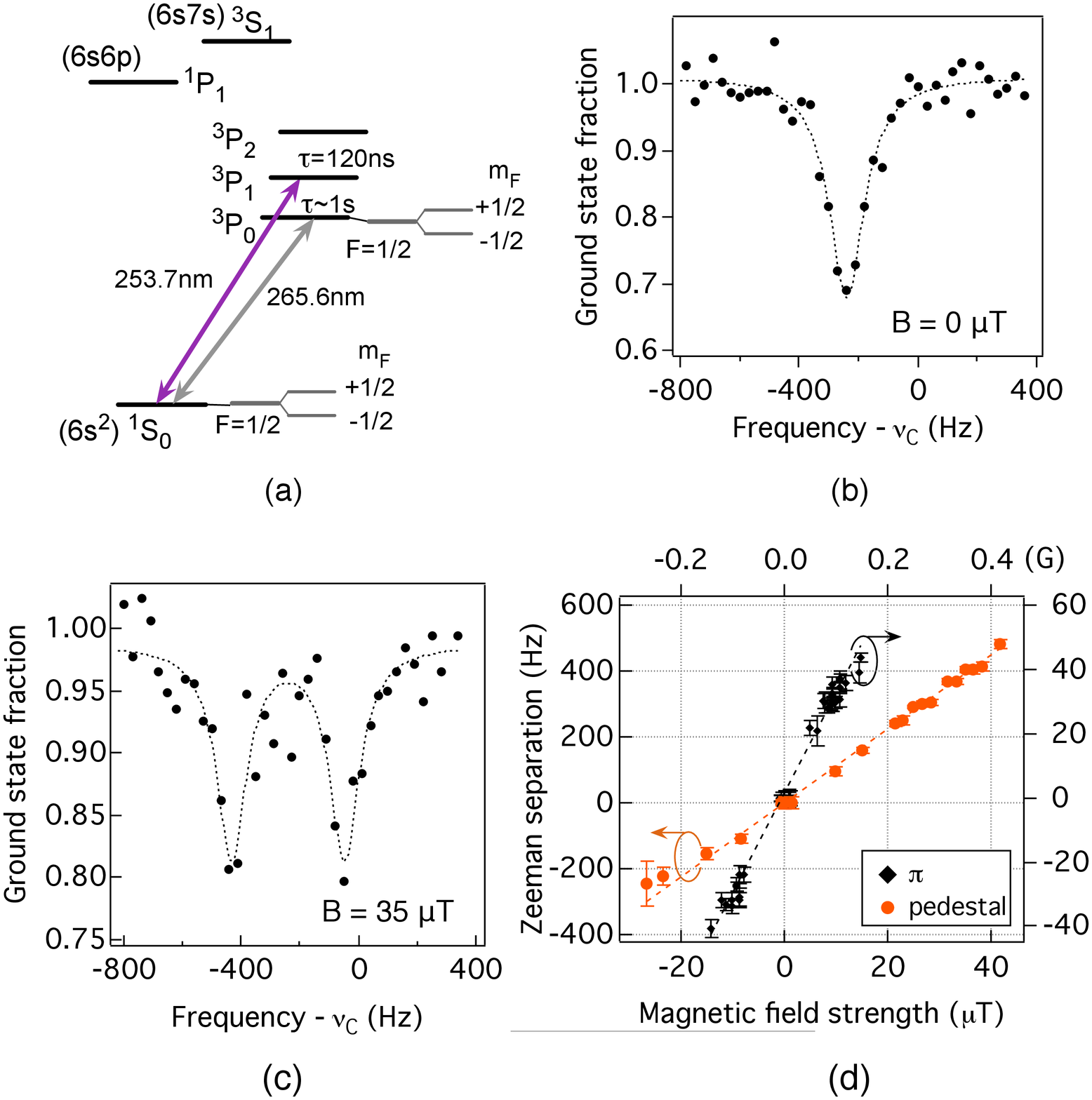}}
\caption[]{\footnotesize (Color online) (a) Partial level scheme for $^{199}$Hg with hyperfine splitting of the ground and excited states. The 265.6\,nm radiation is used to probe the \clockTboth\ clock transition, while 253.7\,nm radiation  is used for cooling and detection. (b) A spectrum of the $^{199}$Hg clock transition ``pedestal'' with m$_{F}=\pm1/2\rightarrow$ m$_{F}=\pm1/2$ Zeeman components overlapped. \lw=362.573\,nm and FWHM = 140\,Hz. (c) Line profile of the $^{199}$Hg pedestal showing the separated Zeeman components with FWHM =120\,Hz. (d) Frequency separation of the Zeeman components for the ultranarrow $\pi$ transition and  pedestal versus bias $B$-field. The slope is $3.1\pm0.3 $\,Hz\,$\mu$T$^{-1}$ and  $11.1\pm1.7$\,Hz\,$\mu$T$^{-1}$, respectively.
} \label{lineprofile}		
\end{center}
\end{figure}

The \clockTboth\ clock transition lies at 265.6\,nm, so as in the case for the 253.7\,nm light generation, two frequency doubling stages are employed (see Fig.~\ref{lineprofile}(a) for the relevant electronic  transitions). The infrared-light (IR) is sourced from a distributed feedback semiconductor laser, injection locked with light from a fiber laser tightly locked to an ultrastable optical cavity~\cite{Daw2010}. About 1\,mW  of 265.6\,nm light is produced by the frequency quadrupling scheme. For rapid control of the 265.6\,nm probe light level we employ an AOM, which positively shifts the frequency of the light by 180\,MHz. Sweeping the frequency of the 265.6\,nm light is described in \cite{Mej2011}. An AOM that is used to suppress noise in the fiber link between the ultrastable laser and the main Hg apparatus is used to tune the clock probe frequency. Despite the drift rate of the ultrastable laser remaining below +30\,mHz\,s$^{-1}$ over the last five months, we still find it helpful to include a de-drift scheme to keep track of the clock transition. This is performed using a direct digital synthesizer (DDS) that steers  the frequency of the AOM in the 1062\,nm path.

The profile of the clock transition is made via detection of atoms in the ground state only and with the timing sequence shown in Fig.~\ref{ChamberandSequence}(b). A broad scan of the transition spectrum is shown in Fig.~\ref{lineprofile}(b), where the magnetic bias field is made small enough that the ($^{1}S_{0}$) m$_{F}=\pm1/2\leftrightarrow$ ($^{3}P_{0}$) m$_{F}=\pm1/2$ Zeeman components overlap one another. The frequency is offset by the value reported in~\cite{Pet2008} for the $^{199}$Hg transition frequency; i.e., $\nu_{c}= 1 128 575 290 808 400$\,Hz. We applied 1.5\,\uW\ of 265.6\,nm light with a e$^{-2}$ beam radius of 310\,\microns\  (intensity = 10\,W\,m$^{-2}$). The \fwhm\ (FWHM) is 140\,Hz and the contrast \si32\,\%.
We show below that a much narrower transition lies at the center of this  spectrum. We will henceforth refer to this broad profile as the pedestal.     This  pedestal may be an indication that  the transverse confinement of the atoms should be improved. When applying a dc $B$-field,  this line profile separates into two Zeeman components [Fig.~\ref{lineprofile}(c)] with slightly lower contrast and FWHM equal to 120\,Hz. The Zeeman line separation versus bias field strength is displayed in Fig.~\ref{lineprofile}(d) (circles) exhibiting a slope of 11\,Hz\,$\mu$T$^{-1}$. Since the 265.6\,nm probe light co-propagates with the lattice light, its  polarization  lies orthogonal to the axial direction of the lattice (i.e. in the horizontal plane). The bias $B$-field is applied in the $x$-direction seen in Fig.\ref{ChamberandSequence}(a). The probe light polarization was confirmed to be mostly linear, but shares a component in both  $x$ and $y$-directions, allowing the possibility for both $\pi$ and $\sigma$ transitions. The 11\,Hz\,$\mu$T$^{-1}$ dependence corresponds closely to that expected for $\sigma$ transitions~\cite{Pet2009a}. Ideally the  probe light polarization should be in the $x$-direction, but in the present experiment the ultranarrow clock line strength (below) is optimised when the polarization has shared $x$ and $y$ components. This is still a matter for investigation.

Information about the confinement of the atoms in lattice trap is garnered by examining the sideband spectra of the Lamb-Dicke spectrum. Although not shown here, when we curve-fit  to the blue sideband using the approach outlined in~\cite{Let2007, Lud2006} with the only free parameters being the lattice depth and the  temperature in the transverse direction, we find $U_{o}=18$\,$E_{R}$ (or 6.5\,\uK) and $T_{r}=7$\,\uK, with associated axial and transverse frequencies of 64\,kHz and 43\,Hz, respectively. We  estimate the temperature of the atoms in the axial direction based on the ratio of the blue and red sidebands to be \si 4\,\uK\ (with $P_{n}/P_{n+1}\sim2.4$). 

\begin{figure}[h]
\begin{center}
{		
\includegraphics[width=8.6cm,keepaspectratio=true]{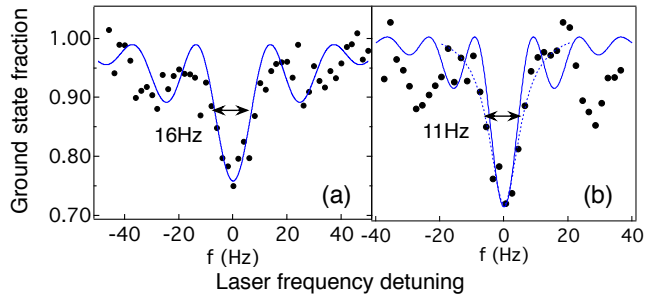}}
\caption[]{\footnotesize(Color online) Ground state fraction versus probe laser detuning frequency for: (a) 50\,ms ($\Omega.T_{p}=1.45 \pi$\,rad) and  (b) 80\,ms  ($\Omega.T_{p}=1.36 \pi$\,rad) Rabi pulses. $B$=0 and \lw$\approx$\mw\ for (a) and (b). 
} \label{UltranarrowSpectrum}		
\end{center}
\end{figure}

With the lattice light set at the magic wavelength  (discussion below) and a higher resolution scan made across the center of the pedestal, we find a much narrower spectral line. Figures~\ref{UltranarrowSpectrum}(a) and (b) show spectra obtained with $T_{p}=50$\,ms and 80\,ms probe pulse durations  taken with approximately 1\,\uW\ and 0.5\,\uW\ of 265.6\,nm probe light, respectively. The solid lines are the modelled Rabi spectra with  $\Omega.T_{p}\sim1.4 \pi$\,rad for both. In Fig.~\ref{UltranarrowSpectrum}(a) $\Omega.T_{p}$ was chosen in an attempt to match to the sidebands observed. While the width of the theoretical trace matches the data, the frequency of the sidebands in general does not, as seen in  \ref{UltranarrowSpectrum}(b). From a measurement of the probe intensity we estimate the Rabi angle to be $\Omega.T_{p} \sim4.5\pi$\,rad. Despite this, we suspect that the sidebands are of a technical origin and are not due to an overdriven Rabi pulse, since reducing the probe power does not decrease the size of the sidebands.  
For most line-center measurements here we use a 50\,ms probe. With an applied bias $B$-field the narrow transition separates into two Zeeman components. The 1$^{\mathrm{st}}$ order Zeeman  dependence is seen in Fig.~\ref{lineprofile}(d) with a slope of  $3.1\pm0.3 $\,Hz\,$\mu$T$^{-1}$.  The ratio of the $B$-field dependencies for the pedestal versus the narrow $\pi$ transition is \si3.6, which is close to the $\sigma/\pi$ ratio of 3.3 that one expects when calculated from the difference in the Land\'{e} $g$-factors of the ground and excited states~\cite{Pet2009a}.

\begin{figure}[h]
\begin{center}
{		
\includegraphics[width=8.6cm,keepaspectratio=true]{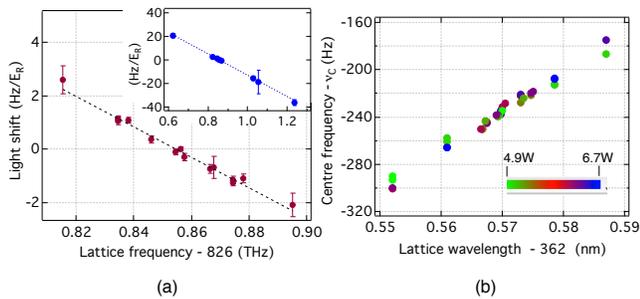}} 
\caption[]{\footnotesize  (Color online) (a) Differential light shift versus lattice frequency for the $^{199}$Hg clock transition.  The inset  shows measurements made over a larger frequency range. The Stark-shift free frequency (wavelength) is  $826.8546\pm0.0024$\,THz ($362.5697\pm0.0011$\,nm)  and the slope is  $-57$\,mHz\,$E_{R}^{-1}$\,GHz$^{-1}$. (b) Clock transition frequency versus lattice wavelength with changes of lattice depth.
} \label{Magicwavelength}		
\end{center}
\end{figure}

For assessments of the ac Stark shift, the center frequency of the clock transition is measured at a series of lattice wavelengths and lattice depths. The results are summarized in Fig.~\ref{Magicwavelength}(a). The main graph shows data  taken with the 10-15\,Hz wide spectral lines, while the inset shows lower resolution data obtained with the pedestal. From these data we determine the magic frequency to be $826.8546\pm0.0024$\,THz  ($\lambda_{m}=362.5697\pm0.0011$\,nm). The uncertainty also incorporates the accuracy of the wavemeter used for the measurements. From the line fit we find the strength of the light shift is  $(-5.1\pm0.9)$\ee{-17}$E_{R}^{-1}$\,GHz$^{-1}$.
Another representation of the light shift is shown in Fig.~\ref{Magicwavelength}(b). Here the line center frequency is plotted as a function of the lattice wavelength while some variation was applied to the lattice depth as indicated by the color scale. The small variation in center frequency near 362.570\,nm is consistent with the above  magic wavelength determination. Here the strength of the light shift is  \si$-6$\ee{-17}$E_{R}^{-1}$\,GHz$^{-1}$.

A preliminary assessment of other systematic shifts affecting the $^{199}$Hg clock transition has been conducted and is summarized in Table~\ref{Budget}. For a temperature uncertainty of 2\,K of the chamber surrounding the atoms, and a 100\,\% uncertainty for the calculated BBR coefficient~\cite{Hac2008}, the shift at 290\,K is $-0.17\pm0.20$\,Hz. A measurement of line center frequency versus Zeeman component separation  showed no variation  within a statistical uncertainty of 1.6\,Hz. This is expected since the predicted 2$\mathrm{^{nd}}$ order Zeeman dependence is 2.16\ee{-9} T$^{-2}$ (e.g., $B=10$\,$\mu$T produces a shift of 0.24\,mHz)~\cite{Hac2008}.

\begin{figure}[h]
\begin{center}
{		
\includegraphics[width=8.5cm,keepaspectratio=true]{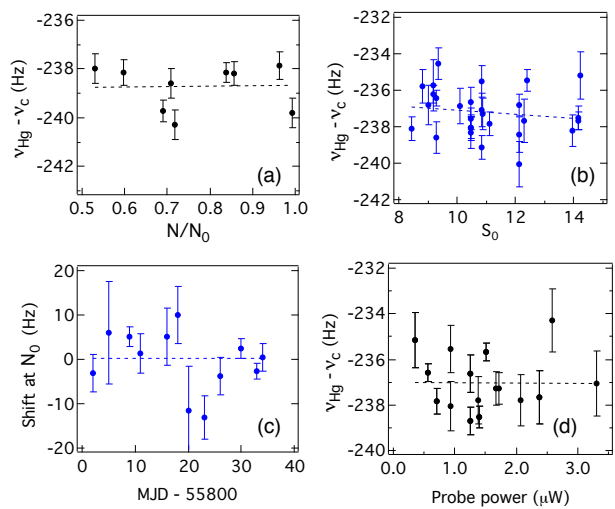}}
\caption[]{\footnotesize   Line center frequency versus: (a) relative atom number, where $N_{0}=2.5$\ee{3} and (b) MOT cooling light intensity, $s_{0}$. (c) Frequency shift at $N_{0}$ versus measurement day. (d) Line center frequency versus probe power. 
} \label{DensityShift}		
\end{center}
\end{figure}

The frequency dependence on atom number has been tested by two means: (i) by varying the loading time of the MOT and (ii) by varying the level of 253.7\,nm light used for cooling. Both methods change the number of atoms loaded into the lattice trap (and also the density since the size of the MOT cloud changes by less than 5\,\% for the range considered here). The nominal atom number is $N_{0}=2.5$\ee{3} (close to the maximum that we achieve). Fig.~\ref{DensityShift}(a) shows the line-center frequency, $\nu_{\mathrm{Hg}}$, versus the relative atom number ($N/N_{0}$) obtained by varying the MOT loading time from 0.8 to 2.5\,s. Fig.~\ref{DensityShift}(b)  shows $\nu_{\mathrm{Hg}}$ for a range of cooling light intensities with a slope equal to $-0.11\pm0.08$\,Hz per unit $s_{0}$ ($s_{0}$ = $I/I_{0}$, where $I_{0}$=102\,W m$^{-2}$). The two results suggest a non-significant density shift at the present resolution. For daily assessments of line-center frequency (described below) there are between 10 and 30 spectra recorded. From a plot of center frequency versus  relative atom number we determine the frequency shift at $N_{0}$, which we show in Fig.~\ref{DensityShift}(c) for each measurement day where there was sufficient variation in $N/N_{0}$. From the weighted mean the density shift  is constrained to $0.26\pm1.9$\,Hz, consistent with the previous two results. Note, the  number of atoms per site is of order one and in the MOT peak atomic density is $\sim6$\ee{10}\,cm$^{-3}$.

An ac Stark shift may also arise due to the probe light. In Fig.~\ref{DensityShift}(d) we show the clock transition frequency versus probe power. At our nominal power level of 1\,$\mu$W the shift lies within the uncertainty of the measurement. We also calculate that the probe light shift is in the $1-10$\,mHz range. Tensor Stark and hyperpolarizability  shifts due to the lattice light are omitted from the table as they are expected to be at least an order of magnitude smaller.

\begin{table}
\caption{Corrections and uncertainties for the $^{199}$Hg clock transition \label{Budget}}
\begin{ruledtabular}
\begin{tabular}{lll}
Effect             & Correction (Hz) & Uncertainty (Hz)  \\
\hline
Blackbody radiation          & -0.17 & 0.2 \\
1st order Zeeman             & $<1$  & 0.5 \\
2nd order Zeeman             & $<1$  & 1.6 \\
Scalar light shift (lattice) & $<1$  & 5.9 \\
Probe light	                 & 0.02  & 0.3 \\
Atom number density		     & 0.4   & 1.9 \\
\hline
Total                        & 0.2   & 6.4
\end{tabular}
\end{ruledtabular}
\end{table}

\begin{figure}[h]
\begin{center}
{		
\includegraphics[width=8.0cm,keepaspectratio=true]{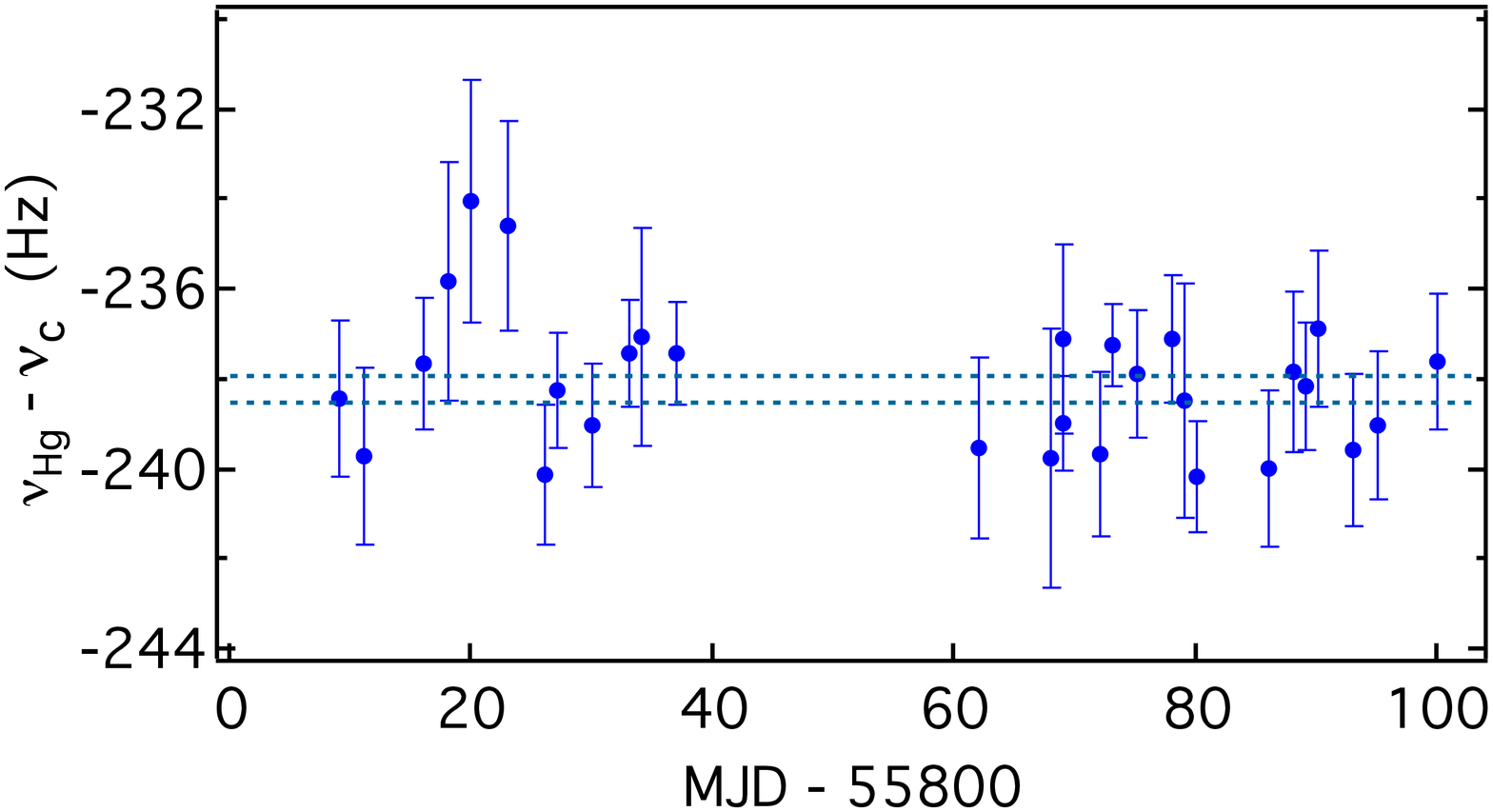}}
\caption[]{\footnotesize  (Color online) Clock transition frequency measured  over a three month period with respect to the SYRTE primary frequency standard for $\lambda_{m} = 362.5697$\,nm. The weighted mean is $\nu_{\mathrm{c}}- 238.2\pm0.3$\,Hz.  $\nu_{\mathrm{c}}= 1 128 575 290 808 400$\,Hz.
} \label{ClockFreqversusTime}		
\end{center}
\end{figure}

Frequency comb accuracy has been verified by comparing measurements from two frequency combs (a Ti:sapphire comb and a Er:fiber based comb), which are steered by a low noise H-maser reference. The two combs show agreement to below 0.01\,Hz. The H-maser reference is continually compared to the LNE-SYRTE  frequency standard, which has a relative uncertainty of 2.7\ee{-16}~\cite{Gue2011b}. An example of the reproducibility of the clock transition frequency is shown in Fig.~\ref{ClockFreqversusTime}, where measurements have been made over a period of three months. Each point is the mean line center frequency produced from between 10 and 30 spectra recorded on each day represented. The measurements were mostly performed with the two ($\pi$-transition) Zeeman components overlapped due to the low S/N in the present experiment. The variations in frequency reduce after MJD = 55825 when we began controlling the lattice light frequency to within 300\,MHz. The uncertainty of the weighted mean of this series is 0.3\,Hz  ($\sigma/\nu_{\mathrm{Hg}}=2.5\times10^{-16}$). Accounting for the systematic shifts (Table~\ref{Budget}) we evaluate the \Hg\ \clockTboth\ transition frequency to be 1 128 575 290 808 162.0 $\pm$ 6.4 (sys.) $\pm$ 0.3 (stat.)\,Hz, where the combined fractional uncertainty is 5.7\ee{-15}.

This work establishes the potential of $^{199}$Hg as a high-accuracy clock. There remain various means for further gains in accuracy; e.g., the S/N of the clock signal should improve as techniques for sideband cooling, transverse cooling and atom number normalization are implemented. One also expects that a reliable means of producing $>150$\,mW of 253.7\,nm radiation will be found in which 2D-MOT loading should increase atom numbers 10-fold~\cite{Pet2009a}.

The authors thank the Syst\`{e}mes de R\'{e}f\'{e}rence Temps-Espace technicians, in particular M. Lours and F. Cornu; G. Santarelli for the use of lab equipment and D.~Magalh\~{a}es for assistance. This work is partly funded by IFRAF and CNES.

\end{document}